\documentclass[manuscript,screen]{acmart} 
\usepackage{graphicx}
\usepackage{booktabs}
\usepackage{float}
\usepackage{mwe}

\title{AR for Sexual Violence: Maintaining Ethical Balance While Enhancing Empathy}

\begin{document}

\author{CHUNWEI LIN}
\email{a108130056@gm2.usc.edu.tw}
\affiliation{%
  \institution{Shih Chien University}
  \city{Taipei}
  \country{Taiwan}
}

\begin{abstract}
 This study showcases an Augmented Reality (AR) experience designed to promote gender justice and increase awareness of sexual violence in Taiwan. By leveraging AR, this project overcomes the limitations of offline exhibitions on social issues by motivating the public to participate and enhancing their willingness to delve into the topic. The discussion explores how direct exposure to sexual violence can induce negative emotions and secondary trauma among users. It also suggests strategies for using AR to alleviate such issues, particularly by avoiding simulations of actual incidents.
\end{abstract}


  
\maketitle
\section{Introduction}
In Taiwan, a prevailing belief holds victims partly responsible for their ordeal, casting doubt on their status as "true" victims. This stigmatization creates an unfriendly social atmosphere, further discouraging victims from seeking external assistance. In response to this pressing issue, the Modern Women's Foundation, a Taiwanese non-governmental organization (NGO) dedicated to advocating for gender justice, conducted Taiwan's first-ever online survey in 2023 to gauge attitudes toward seeking help for sexual violence \cite{MWF_2023}. The survey garnered 876 valid responses, revealing that a third of the respondents (33.4\%, 293 individuals) had experienced sexual violence. Alarmingly, among these respondents, a significant portion (39.2\%) admitted to never disclosing the incident to anyone. The reasons for this silence include (1) the shame associated with the incident, (2) the silence of those around who are aware, (3) the persistent threat from the perpetrator, and (4) the fear of secondary victimization during the disclosure process. To address the critical issue of sexual violence and foster public support for victims, the foundation has traditionally utilized offline exhibitions to educate and engage the public. Recognizing the limitations of traditional methods, the foundation decided to adopt Augmented Reality (AR) in 2023, leveraging AR experience's unique capabilities to enhance public awareness and empathy toward social justice issues.

In the past, engaging the public in offline exhibitions on social justice topics such as sexual violence has been challenging. While motivation is a primary factor in participating \cite{Cheng_Xu_Pan_2023}, individuals may be less inclined to engage with these exhibitions, perceiving them as irrelevant to their life experiences and offering fewer attractions than art or entertainment exhibitions. By adopting AR experiences, interactive technology features not only boost public engagement \cite{Chang_Horng_2010,Cheng_Xu_Pan_2023} but its capacity to establish an Immersive Virtual Environment (IVE) \cite{Usoh_Slater_1995} also cultivates a sense of presence \cite{Lombard_Ditton_2006}. This feeling of "being there" is essential in fostering intergroup empathy \cite{Batson_Ahmad_2009,Sundar_Kang_Oprean_2017,Goutier_2021}, thus rendering the content of social justice engaging and encouraging individuals to delve further into the topic \cite{Pjesivac_Ahn_Briscoe_Kim_2022}. Such an approach offers a good starting point for understanding the exhibition's content. Moreover, integrating AR experiences with offline exhibitions is crucial since the empathy evoked through IVE primarily focuses on emotional empathy, which enables participants to feel the emotions of others physically \cite{Martingano_Hererra_Konrath_2021}. While this form of empathy is impactful, it lacks the inclusion of cognitive empathy, which is essential for grasping the perspectives and thoughts of those affected \cite{Davis_1983,Decety_Jackson_2004}. Therefore, combining AR with offline exhibition ensures a holistic educational approach, allowing the public to not only feel but also understand the complex situations faced by victims of sexual violence.

\section{Implementation}

\begin{figure}[H]
\centering
\includegraphics[width=\textwidth]{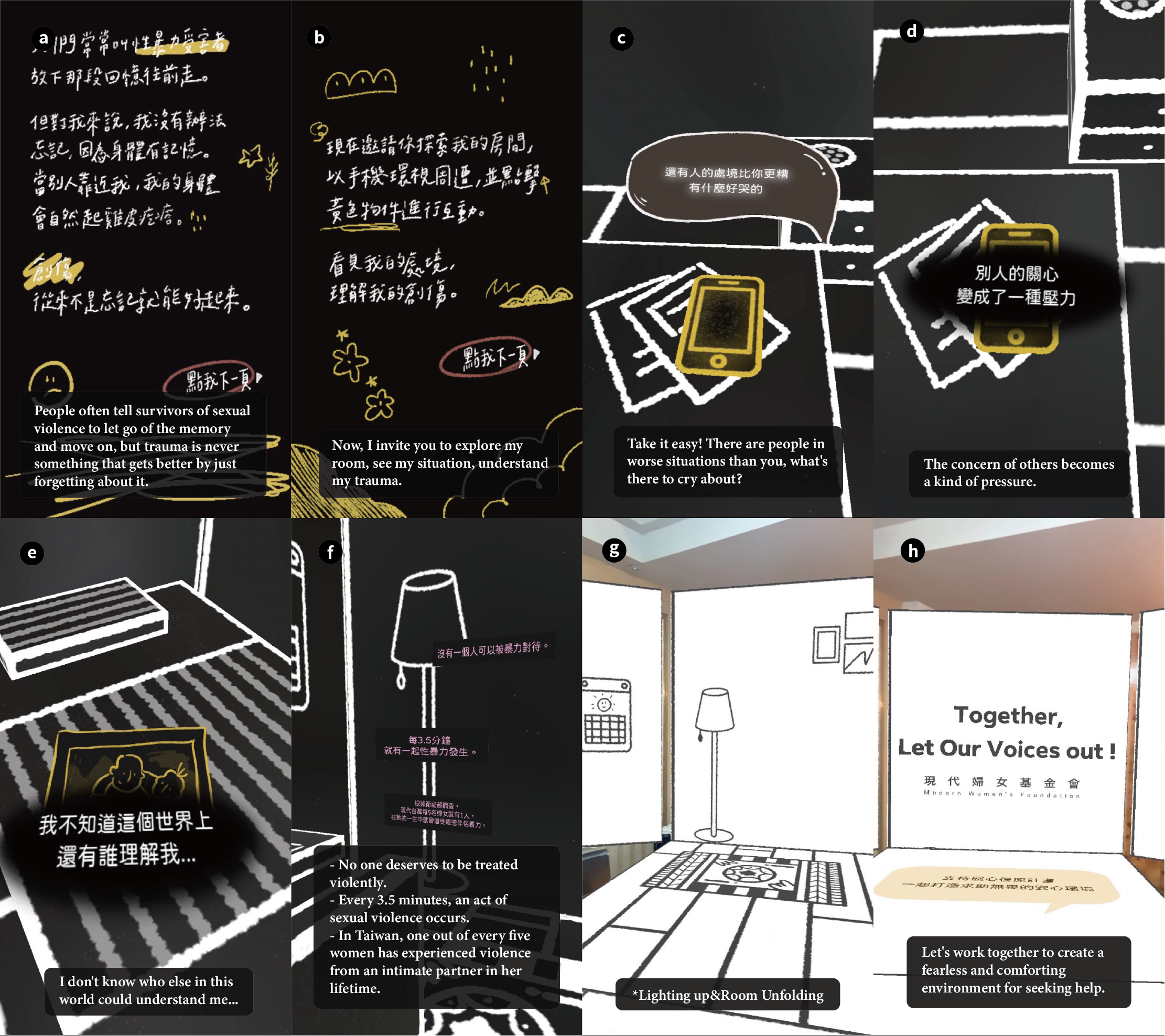}
\caption{Screenshots of the "\textit{Light up the Room}" project. (a-b) Initial instruction prompt. (c-f) Exploration process inside the victim's room. (g-h) Appearance of the room once illuminated and unfolded.}
\label{Fig1}
\Description{Screenshots of the Light up the Room project}
\end{figure}

\textit{Light up the Room}, an AR experience created by Pei-Chi Wang, Han-Yun Chung, and Jia-Zhu Chu, was designed to encourage public engagement with issues of sexual violence and to warm the hearts of those affected by the tragedy through engaging interactions that cast light into the survivors' shadowed spaces. For a closer look, please watch the full video \cite{Wang_Chu_Chung_2024} or visit Instagram to experience it \cite{Filter}.

To interact with the artwork, users are introduced to a survivor's internal monologue (Figures \ref{Fig1}a-b), followed by instructions to explore the victim's room. Here, tapping on objects reveals the challenges faced by sexual violence survivors (Figure \ref{Fig1}c); the corresponding interactive events to the challenges are presented in Table \ref{events}. Interacting with these objects not only changes their color but also displays messages of the victim's inner thoughts or statistics relevant to sexual violence (Figures \ref{Fig1}d-f), ultimately transforming the room from dark to light. This gives the appearance that the user has genuinely illuminated the room (Figure \ref{Fig1}g). Then, the walls rotate outward, opening the space to reveal the real-world environment outside the previously enclosed space (Figure \ref{Fig1}g). This interactive journey aims to deepen public understanding of survivors' experiences and create a more comforting environment for them to seek external support.

\begin{table}[H]
\resizebox{\textwidth}{!}{%
\begin{tabular}{@{}lll@{}}
\toprule
\textbf{Situation} & \textbf{Object} & \textbf{Event}       \\ \midrule
Distrust of others & Group photo     & A photo torn in half \\[10pt]
Self-doubt, self-scrutiny                  & Mirror     & A voiceover plays, revealing the victim’s internal thoughts while looking in the mirror \\[10pt]
Secondary victimization by internet trolls & Computer   & Comments from trolls continuously emerge from the screen                                \\[10pt]
Lack of understanding from family          & Smartphone & Displays misguided concern from family members rising from the phone                    \\[10pt] \bottomrule
\end{tabular}%
}
\caption{Corresponding interactive events to the challenges}
\label{events}
\end{table}

In the platform selection process, Meta Spark AR on Instagram was selected for several reasons: (1) For ease of access, Instagram boasts a substantial user base in Taiwan, with 10.87 million individuals, or 47\% of the population, actively engaging on the platform as of January 2023 \cite{OOSGA_2023}. This accessibility makes it a low-friction option for reaching audiences. (2) From a development perspective, Meta Spark Studio offers an intuitive user interface that facilitates real-time feedback on modifications, which is essential for efficiently refining AR effects. Its 'patch editor,' a node-based editing system, simplifies the development process, enabling creators to produce content more swiftly and effortlessly. (3) Considering the cost, Meta Spark Studio is available at no charge, which is particularly beneficial for NGOs operating on tight budgets. These factors combined make Meta Spark Studio the optimal AR development platform for Light up the Room. According to Meta Spark Policies \cite{Meta_spark_policies}, two potential drawbacks must be considered despite the advantages: (1) To ensure suitability for the general user base, AR content must not be shocking, sensational, disrespectful, or violent. This gives the platform interpretive space to refuse content that they consider inappropriate. (2) External links and tags within filters, including URLs, QR codes, or other scannable codes, are prohibited. This could limit the promotion of social justice issues, preventing users from accessing further information. Regarding user data privacy, according to Instagram's Privacy Policy \cite{Instagram_policies}, Meta collects data, including users' AR usage behavior; however, this data is intended for personalized experiences and advertising purposes.

\section{Discussion}

\begin{figure}[!htp]
\centering
\includegraphics[width=0.48\linewidth]{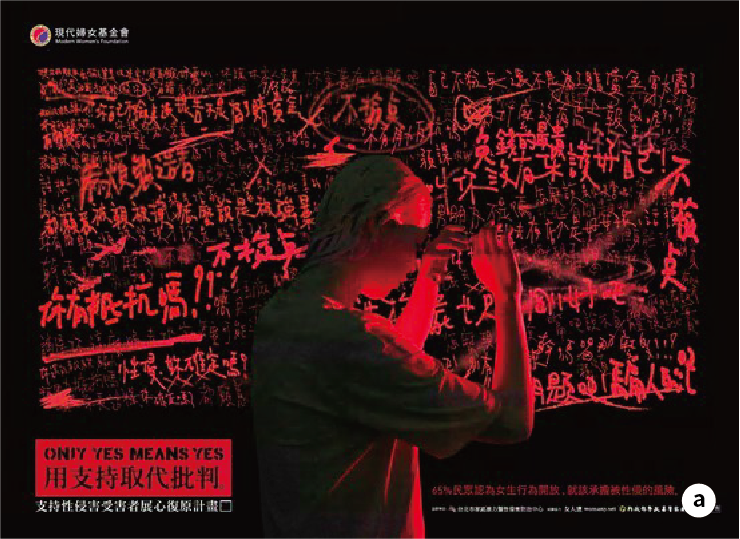}
\Description{MRT lightbox}
\hfill
\includegraphics[width=0.48\linewidth]{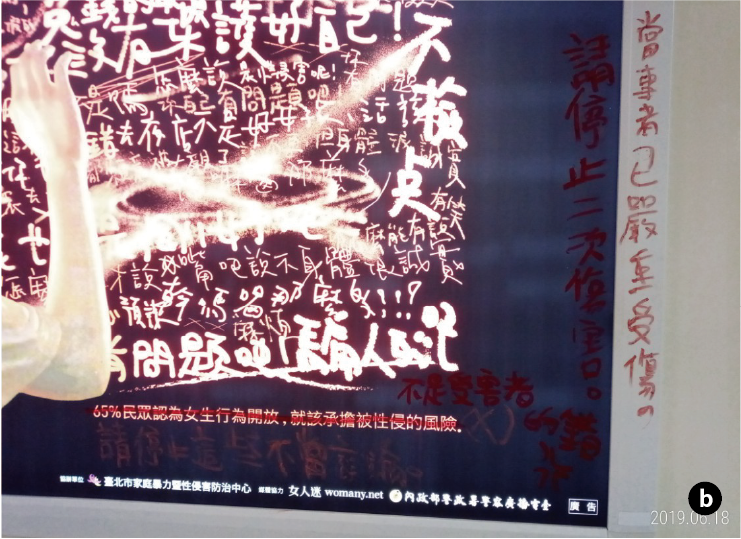}
\Description{Protest messages}
\caption{(a) Lightbox advertisement placed by the Modern Women's Foundation in Taipei Metro in 2019. (b) Protest messages written by victims in lipstick alongside the advertisement.}
\label{Fig2}
\end{figure}

\textit{Light up the Room} simulates the environment of a survivor's room after the incident, deliberately shifting the focus away from the actual moment of sexual violence. This approach stems from the Modern Women's Foundation's "Replace Criticism with Support" campaign in 2019, which aimed to highlight the harm of negative remarks that oppressed sexual violence victims, using a lightbox ad in the Taipei Metro (Figure \ref{Fig2}a). However, a victim misinterpreted the ad, reacting with messages of protest written in lipstick, such as "It's not the victim's fault," "The victim has been severely harmed," and "Please stop these inappropriate remarks" (Figure \ref{Fig2}b). This unintended effect, which led to the ad's removal, showcased the complexity of addressing sexual violence publicly and the potential for secondary trauma, prompting a rethink on the approach to such sensitive issues. 

The goal of the AR experience is to promote an understanding of the victims' circumstances. Yet, while AR can deepen empathy by creating a sense of presence, direct experiences of sexual violence in immersive virtual environments (IVE) may elicit negative emotions and anxiety among both users and victims \cite{Loranger_Bouchard_2017}, which may be inappropriate for an advocacy-oriented scenario and could inadvertently spark controversy. It risks causing secondary trauma to survivors and may also induce anxiety and distress in users. Avoiding the depiction of the incident itself can mitigate the discomfort experienced by survivors. Hence, by deliberately shifting the focus away from the actual moment of sexual violence, \textit{Light up the Room} aimed to address the aftermath without exacerbating the trauma experienced by survivors. In practical application, it was met with positive responses for creating an immersive experience and its educational value in conveying the experiences of victims. Nonetheless, a few survivors reported discomfort, underscoring the importance of finding a balance between educational content and emotional sensitivity. This feedback underscores the imperative for additional research to perfect this equilibrium and ensure that these tools are potent and considerate in advocacy work.

\section{Conclusion}
While it is widely believed that AR, as a medium, can enhance understanding of issues and inspire action through immersive experiences, direct exposure to sexual violence is not the most effective approach for the success of advocacy-oriented AR experiences. This paper seeks to offer insights for future AR projects in social justice, guiding them toward a finely balanced user experience that navigates between empathetic engagement and impactful action.

\begin{acks}
Thanks go to Pei-Chi Wang, Han-Yun Chung, and Jia-Zhu Chu, the creators of Light up the Room, as well as Jing-Yuan Huang for their contributions to data collection and their valuable feedback on an early draft of this work. Gratitude is also extended to the Modern Women's Foundation for their collaboration.
\end{acks}

\bibliographystyle{ACM-Reference-Format}
\bibliography{ref}

\end{document}